# The role of primary point defects in the degradation of silicon detectors due to hadron and lepton irradiation[&]

**I. Lazanu**

University of Bucharest, Faculty of Physics, POBox MG-11, Bucharest-Magurele, Romania
e-mail: i_lazanu@yahoo.co.uk

**S. Lazanu** [*]
National Institute for Materials Physics, POBox MG-7, Bucharest-Magurele, Romania,
e-mail: lazanu@infim.ro

**Abstract**

The principal obstacle to long-time operation of silicon detectors at the highest energies in the next generation of experiments arises from bulk displacement damage which causes significant degradation of their macroscopic properties. The analysis of the behaviour of silicon detectors after irradiation conduces to a good or reasonable agreement between theoretical calculations and experimental data for the time evolution of the leakage current and effective carrier concentration after lepton and gamma irradiation and large discrepancies after hadron irradiation and this in conditions where a reasonable agreement is obtained between experimental and calculated concentrations of complex defects.
In this contribution, we argue that the main discrepancies could be solved naturally considering as primary defects the self-interstitials, classical vacancies and the new predicted fourfold coordinated silicon pseudo-vacancy defects. This new defect is supposed to be introduced uniformly in the bulk during irradiation, has deep energy level(s) in the gap and it is stable in time. Considering the mechanisms of production of defects and their kinetics, it was possible to determine indirectly the characteristics of the $Si_{FFCD}$ defect: energy level in the band gap and cross section for minority carrier capture. In the frame of the model, the effects of primary defects on the degradation of silicon detectors are important in conditions of continuous long time irradiation and /or high fluences.

---

# 1. Introduction

Particle physics makes its greatest advances with experiments at the highest energies. The way to advance to an energy regime higher than the actual one is through a new generation of colliders or through non-accelerator experiments. In the near future, the Large Hadron Collider (LHC) will be operational, and beyond that, its upgrades: the Super-LHC (SLHC) and the hypothetical Very Large Hadron Collider (VLHC), or International Linear Collider (ILC). AMS, Pamela or GLAST for example will represent new astroparticle physics experiments.

Silicon detectors will be used extensively in the next generation of experiments in high energy physics, where they will operate in a high radiation background due to high fluences of hadrons and leptons.

When an incident particle is slowed down in silicon, it produces different types of damage. They depend on the competition between the cross sections of the corresponding processes, e.g. energy transfer to atomic electrons (ionisation) and energy transfer to translational motion of the atom as a whole. While ionisation is the basis of particle detection and is reversible, displacement effects produce defects in the lattice, determining changes in the material and device characteristics.

Thus, the principal obstacle to long-time operation arises from bulk displacement damage in silicon, which produces primary point defects. During annealing processes, these defects interact between them or with impurities and produce new defects, which increase the leakage current in the detector, decrease the satisfactory Signal/Noise ratio, and increase the effective carrier concentration (and thus the depletion voltage), which ultimately increase the operational voltage of the device beyond the breakdown voltage.These effects must be considered in the design of semiconductor detectors for high energy physics.

In the analysis of the behaviour of silicon detectors after irradiation, the use of a model of defect kinetics and the SRH statistics conduces to a good or reasonable agreement between theoretical calculations and experimental data for the production and time evolution of the leakage current and effective carrier concentration after lepton and gamma irradiation. After hadron irradiation, discrepancies up to 2 orders of magnitude are reported, and this in conditions where a reasonable agreement is obtained between experimental and calculated concentrations of complex defects. Similar discrepancies have been reported by different groups [1, 2] and some explanations are searched, in fact all associated with the possible existence of defects E70, E170, P6 di-interstitials, $V_2$ as clusterised defects [2, 3, 4, 5, 6], contrary to gamma or lepton case where MacEvoy and co-workers claimed that the behaviour of silicon detectors could be explained mainly with the contribution of the $V_2O$ defect and with the consideration of the inter-defect charge exchange mechanism.

In this paper, we argue that the discrepancies mentioned before could be solved naturally if we consider the existence, as primary defects, of self-interstitials, classical vacancies and fourfold coordinated silicon pseudo-vacancy defects, $Si_{FFCD}$, all affecting the characteristics of detectors. Goedecker and co-workers [7] predicted theoretically the existence of this defect, with energy level(s) in the band gap. The energy level in the band gap and cross-section for minority carrier capture are indirectly determined in the present work, imposing the condition that the results of the model must reproduce available experimental data obtained after lepton and hadron irradiation.

In concrete calculations, the kinetics of defects is treated in the frame of a model based on the theory of diffusion limited reactions, developed previously by the authors; see for example [8] and references cited therein.

The model has no free parameters and is able to predict absolute values for leakage current and effective carrier concentration for different silicon materials used for detectors and in various radiation fields.

# 2. Macroscopic characteristics of silicon detectors

The dark current of a reverse biased *p - n* junction is composed of the following terms: the generation current, due to carrier generation on the midgap energy levels inside the depleted region, the drift current, due to the drift of minority carriers and surface and perimetral currents, dependent on the environmental conditions of the surface and the perimeter of the diode.
The increase of the leakage current during and after irradiation is due to the generation of electron-hole pairs on the energy levels of defects produced, and could be evaluated in the frame of the simplified Shockley-Read-Hall model [9], [10].In this model each defect is considered with one or more levels in the bandgap, uncoupled, and thus the current is simply the sum of defect contributions. The defects which have an important contribution to the leakage current are those with energy levels in the vicinity of the mid-gap, and with high cross sections for carrier capture.

Supposing the detector a totally depleted p - n junction, the increase of the volume density of the leakage current, j, could be approximately written in agreement with SRH model as:

$$j = q \langle v_{th} \rangle n_i \left[ \sum_d N_d \sigma_d \exp\left(-\frac{|E_d - E_i|}{kT}\right) + \sum_a N_a \sigma_a \exp\left(-\frac{|E_a - E_i|}{kT}\right) \right] \quad (1)$$

The absolute value of the difference between ionised donors and acceptors in the space charge region of the detector is the effective carrier concentration in the space charge region, $N_{eff}$:

$$N_{eff} = \left| [P] - [VP] + \sum_d \left(\frac{N_C}{N_V}\right) \cdot \left(\frac{\sigma_n}{\sigma_p}\right)_d \cdot \exp\left(-\frac{2|E_d - E_i|}{kT}\right) - \sum_a \left(\frac{N_V}{N_C}\right) \cdot \left(\frac{\sigma_p}{\sigma_n}\right)_a \cdot \exp\left(-\frac{2|E_a - E_i|}{kT}\right) + N_{sd} - N_{sa} \right| \quad (2)$$

In the two equations, index "*d*" is associated with deep donor defects "*a*" with deep acceptors, "sd" and "sa" with shallow donors and acceptors respectively. Here $\sigma_n (\sigma_p)$ are the cross sections for the capture of majority (minority) carriers, $E_i$ is the intrinsic level, $n_i$ the intrinsic concentration of carriers and $\langle v_t \rangle$ is the average between electron and hole thermal velocities, with *q* – the electric charge of the electron. $N_C$ and $N_V$ - the effective densities of states in the conduction (valence) band.

## 3. Modelling of the radiation effects
### 3.1. Aspects associated with primary defects in silicon

The stability of crystalline silicon comes from the fact that each silicon atom can accommodate its four valence electrons in four covalent bonds with its four neighbours. The production of primary defects or the existence of impurities or lattice defects destroys the fourfold coordination.

It has been established that the structural characteristics of the "classical" vacancy are: the bond length in the bulk is 2.35Å and the bond angle – 109°. The formation energy is 3.01 eV (p-type silicon), 3.17 eV (intrinsic), 3.14 eV (n-type).

For interstitials, different structural configurations are possible: a) the hexagonal configuration, a sixfold coordinated defect with bonds of length 2.36 Å, joining it to six neighbours which are fivefold coordinated; b) the tetrahedral interstitial is fourfold coordinated; has bonds of length 2.44 Å joining it to its four neighbours, which are therefore five coordinated; c) the split - <110> configuration: two atoms forming the defect are fourfold coordinated, and two of the surrounding atoms are fivefold coordinated; d) the 'caged' interstitial contains two normal bonds, of length of 2.32 Å, five longer bonds in the range 2.55÷2.82 Å and three unbounded neighbours at 3.10÷3.35 Å. The calculations [11], [12], [13] found that the tetrahedral interstitial and caged interstitial are metastable. For interstitials, the lowest formation energies in eV are 2.80 (for p-type material), 2.98 (for n-type) and 3.31 in the intrinsic case respectively.

It has been established that in silicon the vacancy takes on five different charge states in the band gap: $V^{2+}$, $V^+$, $V^0$, $V^-$, and $V^{2-}$ and the self-interstitial could exists in four charge states after some authors [14]: $I^-$, $I^0$, $I^+$ and $I^{2+}$, or in five states, after other authors [15, 16].

The experimental examination of primary point defects buried in the bulk is difficult and for various defects this is usually indirect. In a series of theoretical studies [17] and correlated EPR and DLTS experiments of Watkins and co-workers [18], it became possible to solve some problems associated with the electrical level structure of the vacancy. The charge states $V^{2+}$, $V^+$, $V^0$ form the so-called negative U system, caused when the energy gain of a Jahn-Teller distortion is larger than the repulsive energy of the electrons, case in which the (0/+) level is inverted in respect to (+/++) level, which are the striking consequence of the fact that the $V^+$ charge state is metastable.

Only recently, Lukjanitsa [14] identified experimentally all the energy levels assigned to vacancies and interstitials. In the following two tables we present a review of the present knowledge on energy levels in the band gap for isolated vacancies and interstitials, as experimental data and model calculation together with the corresponding charge states.

**Table 1:** Energy levels of isolated vacancies in silicon

| Vacancy | | | |
|---|---|---|---|
| Energy level [eV] | | Reference | Assigned charge state |
| Experimental | Calculated | | |
| $E_v + 0.05$ | | [18] | $V^{+/0}$ |
| $E_v + 0.13$ | | [18] | $V^{2+/+}$ |
| | $E_v + 0.36$ | [19] | $V^{0/-}$ |
| $E_v + 0.47$ | | [14] | Non attributed |
| | $E_v + 0.76$ | [20] | Non attributed |
| $E_v + 0.84$ | | [14] | Non attributed |
| | $E_v + 0.84$ | [19] | $V^{2-/-}$ |
| $E_v + 1.01$ | | [21] | $V^{2-/-}$ |

**Table 2**: Energy levels of isolated interstitials in silicon

| Interstitial | | | |
|---|---|---|---|
| Energy level [eV] | | Reference | Assigned charge state |
| Experimental | Calculated | | |
| | $E_v + 0.12$ | [20] | Non attributed |
| $E_v + 0.26$ | | [14] | Non attributed |
| | $E_v + 0.4$ | [16] | $I^{2+/+}$ |
| $E_v + 0.45$ ? | | [14]. | Non attributed |
| | $E_v + 0.52$ | [20] | Non attributed |
| $E_v + 0.68$ | | [14] | Non attributed |
| | $E_v + 0.7$ | [16] | $I^{+/0}$ T-X cross |
| | $E_v + 0.76$ | [15] | $I^{2+/-}$ |
| | $E_v + 0.9$ | [16] | $I^{+/0}$ T-T cross |
| | $E_v + 1.04$ | [15] | $I^{-/2-}$ |

Recently, using state of the art plane wave density functional theory, Goedecker and co-workers [7] predicted the existence of a new type of primary defect: $Si_{FFCD}$ (**F**our**f**olded **C**oordinated silicon **D**efect) that is a pseudo-vacancy. It is obtained by moving atoms from the initial positions, but this displacement does not break the bonds with the neighbours. The bound lengths are between 2.25÷2.47 Å and angles vary in the 97÷116° range. So, the bound length and angle do not deviate significantly from their bulk values. The formation energy is 2.45 eV (for p-type silicon), 2.42 eV (intrinsic), 2.39 eV (n-type), lower than the energy of formation of both vacancies and interstitials. For this defect, it was predicted that it has energy levels in the band gap and most probably it is very stable.

### 3.2. Complex defects and their kinetics. Basic model hypothesis and concrete mechanisms

The model of kinetics of defects used in the present work has been developed previously by the authors [22]. It describes the formation and evolution of defects in silicon during and after irradiation.

In the first step of the model, the incident particle interacts with the semiconductor material. The peculiarities of the interaction mechanisms are explicitly considered for each particle and kinetic energy.

In the second step, the recoil nuclei resulting from these interactions lose their energy in the semiconductor lattice. Their energy partition between displacements and ionisation is considered in agreement with Lindhard's theory [23], [24]. The kinetic energy dependence of the concentration of primary defects on unit particle fluence (CPD) is thus calculated for each incident particle, and it is used in the evaluation of the generation rate of defects during irradiation.

The third step is devoted to the study of formation and time evolution of complex defects, associations of primary defects or of primary defects and impurities. The primary point defects considered here are vacancies (as "classical" vacancies and fourfold coordinated vacancy defect: $Si_{FFCD}$) and silicon self interstitials.

Related to the $Si_{FFCD}$ defect we supposed that this defect is produced simultaneously with the classical vacancy with a concentration that must be determined, it is uniformly introduced in the bulk during irradiation, has deep energy level(s) in the gap, and is stable in time. In these hypotheses, the reaction of production of primary defects is:

$$Si \xrightarrow{G} (V + Si_{FFCD}) + I \qquad (3)$$

The primary defects are created at a rate $G$, sum of contributions from thermal generation and irradiation; the rate of defect production due to irradiation being calculated as:

$$G_{irradiation} = \int CPD(E) \times Flux(E) dE \qquad (4)$$

The energy dependence of the CPD produced by different incident particles could be found, e.g. in Figure 1 of reference [25]. Applying the theory for diffusion-limited reactions, the formation of defects could be described by the reaction scheme indicated in the Table 3. The activation energies of the considered processes are also specified:

**Table 3.** The kinetics of defects

| Reaction | Activation energy [eV] | |
|---|---|---|
| | Direct process | Reverse process |
| $V + I \rightarrow Si$ | 0.352 | |
| $I \rightarrow sinks$ | 0.43 | |
| $V + V \leftrightarrow V_2$ | 0.75 | 1.5 |
| $V + P \leftrightarrow VP$ | 0.7 | 1.37 |
| $V + O \leftrightarrow VO$ | 0.74 | 1.51 |
| $I + C_s \leftrightarrow C_i$ | 0.252 | 1.5 |
| $C_i + C_s \leftrightarrow C_i C_s$ | 0.66 | 1.67 |
| $C_i + O_i \leftrightarrow C_i O_i$ | 0.66 | 1.67 |

In the values used for the activation energies, the contribution of a Coulomb potential between ionised defects is considered. A detailed discussion about these interactions is done by Fahey and co-workers [21].

The time evolution of defect concentrations for primary and complex defects produced in silicon is a solution of the associated system of simultaneous differential equations. This system is solved numerically.

The results of the modelling are imposed to reproduce experimental data simultaneously both at the microscopic (concentration of the defects) and macroscopic level (leakage current and effective concentration of charge carriers).
The fraction of the vacancies that forms the $Si_{FFCD}$ defect, the corresponding energy level in the band gap and the cross section for carrier capture represent the parameters of the model. In these conditions, the characteristics of the new defect are indirectly established.

## 4. Results, discussions and predictions
### 4.1 Microscopic characteristics

For concrete calculations, the experimental data for the time evolution of the concentration for VP, VO and interstitial C defects after electron irradiation from Song's paper [26] have been used. The concentrations of defects have been calculated numerically, as solutions of the system of differential equations associated with the reaction scheme. The reaction constants have been evaluated using data from the literature, and in the case these were not available, adjusted in such a way that they could reproduce the time evolution of defect concentrations measured experimentally. The values of the activation energies of the processes considered are those from Table 3.

In Figure 1, the data for the time evolution of the concentrations of VO, VP and $C_i$ after electron irradiation [26] are represented on the same graph with model calculations.

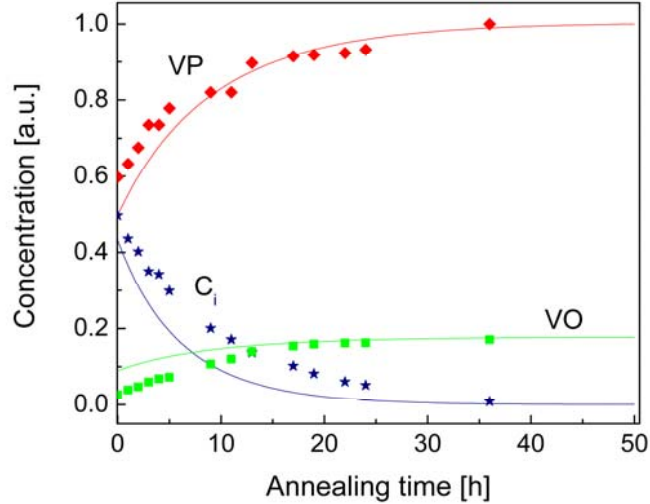

**Figure 1**
**Evolution of defect concentrations in irradiated silicon: experimental data [26] and calculations in the present work**

### 4.2 Macroscopic characteristics

We also calculated the modifications induced by irradiation in the leakage current and effective carrier concentration of silicon detectors, supposing the existence of $Si_{FFCD}$ defect and compared them with available experimental data for the time dependence of the leakage current after electron [27], proton [28], positive and negative pion [29] irradiation; and for the time dependence of the carrier concentration in the space charge region of the detector after proton [28, 30], pion [29] and neutron [31] irradiation.

For the defects other than $Si_{FFCD}$, with relevant contributions to macroscopic modifications of detectors characteristics, the energy levels in the band gap and the cross-sections for minority carriers used in the concrete calculations are indicated in Table 4.

**Table 4**:
Energy levels in the band gap and cross sections for minority carriers used the modelling.

| Defect | $E_T$ [eV] | Cross section [cm$^2$] | Ratio of cross sections |
|---|---|---|---|
| V | $E_v + 0.47$ | $\sigma_p = 9 \cdot 10^{-16}$ | $\sigma_n / \sigma_p = 1$ |
| I | $E_c - 0.44$ | $\sigma_n = 2 \cdot 10^{-15}$ | $\sigma_p / \sigma_n = 50$ |
| VP | $E_c - 0.46$ | $\sigma_n = 4 \cdot 10^{-15}$ | $\sigma_p / \sigma_n = 50$ |
| $V_2$ | $E_c - 0.43$ | $\sigma_n = 2 \cdot 10^{-15}$ | $\sigma_p / \sigma_n = 25$ |
| $C_iO_i$ | $E_v + 0.36$ | $\sigma_p = 2 \cdot 10^{-15}$ | $\sigma_n / \sigma_p = 10$ |

The numerical values utilised in the calculations are in agreement with Refs.[14], [32], [33] and [34].

In figures 2 a, b, c, d, the time dependence of the degradation constant of the leakage current after irradiation with electrons of 800 MeV kinetic energy, protons of 24 GeV/c, negative and positive pions at 350 MeV/c, at room temperature, is represented.

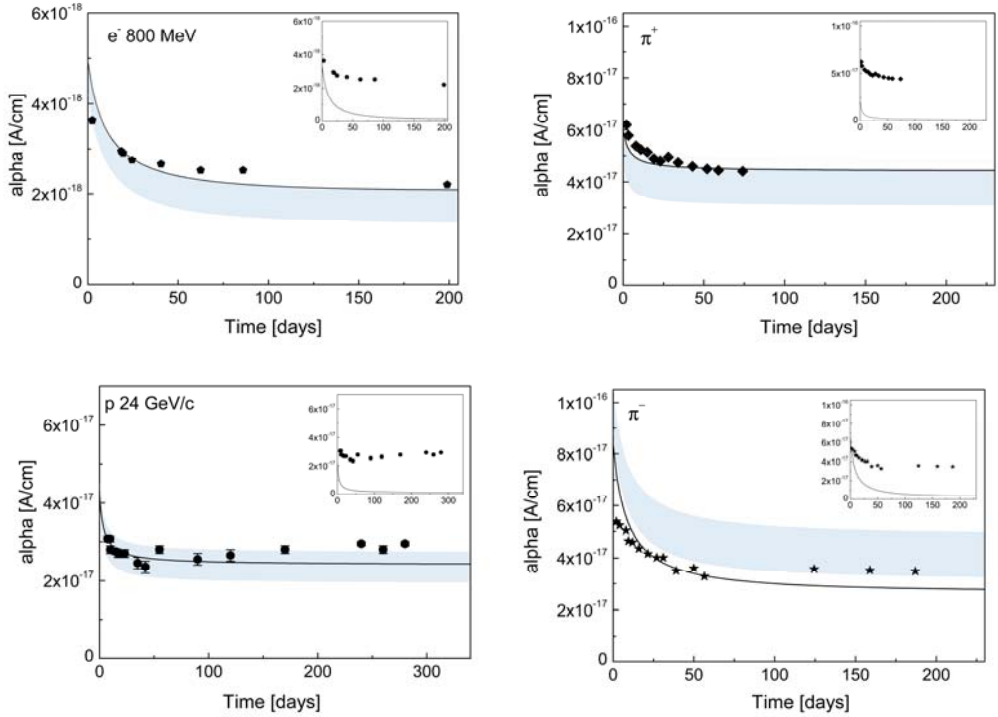

**Figures 2 a ÷ d**
The damage constant of the leakage current as a function of time after irradiation at room temperature, with:
a) electrons of 800 MeV kinetic energy; b) protons of 24 GeV/c momentum; c) and d) negative and positive pions at 350 MeV/c. In the insert: comparison between experimental data (points) and model calculations without the contribution of the $Si_{FFCD}$ defect (continuous line).

In figures 3 a ÷ e the results corresponding to the calculations for the time dependence of the absolute values of effective carrier concentration are given. Figure 3a refers to neutrons, 3b and 3c represent results for protons of 24 GeV/c and 800 MeV respectively, 3d and 3e correspond to negative and positive pions of 350 MeV/c.

In all figures, continuous lines are model calculations with the parameters of the $Si_{FFCD}$ defect optimised to best reproduce the data. The shadow bands represent the interval of values for one standard deviation around the average contribution of the $Si_{FFCD}$ defect over all experimental considered data. The experimental data are represented by symbols. In the insert, experimental data (points) are represented together with model calculations (lines), without the contribution of the $Si_{FFCD}$ defect.

From this analysis, we obtained that the $Si_{FFCD}$ defect is produced with a concentration of about 10% from all vacancies per act of interaction, the defect has an energy level in the band gap between $E_c - (0.46 \div 0.48)$ eV, a capture cross section between $(5 \div 10) \times 10^{-15}$ cm$^2$ and a ratio $\sigma_p / \sigma_n = 1 \div 5$.

Some delayed mechanisms of formation of complex defects which were not considered in the model could also exist, and their contributions become observable long time after irradiation, as the evolution of macroscopic characteristics long time after irradiation suggests. Also, a relatively lower accuracy of the calculated results for $\pi^+$ and $\pi^-$ cases could be due to the fact that the differences between the two particles in the calculation of the CPD is not considered. In spite of the simplicity of the model, the stability of the parameters of the defect determined by this procedure must be remarked.

These results need experimental confirmation. The fact that the $Si_{FFCD}$ defect has not yet been detected experimentally must be explained, and probably a re-examination of different experimental results will be necessary.

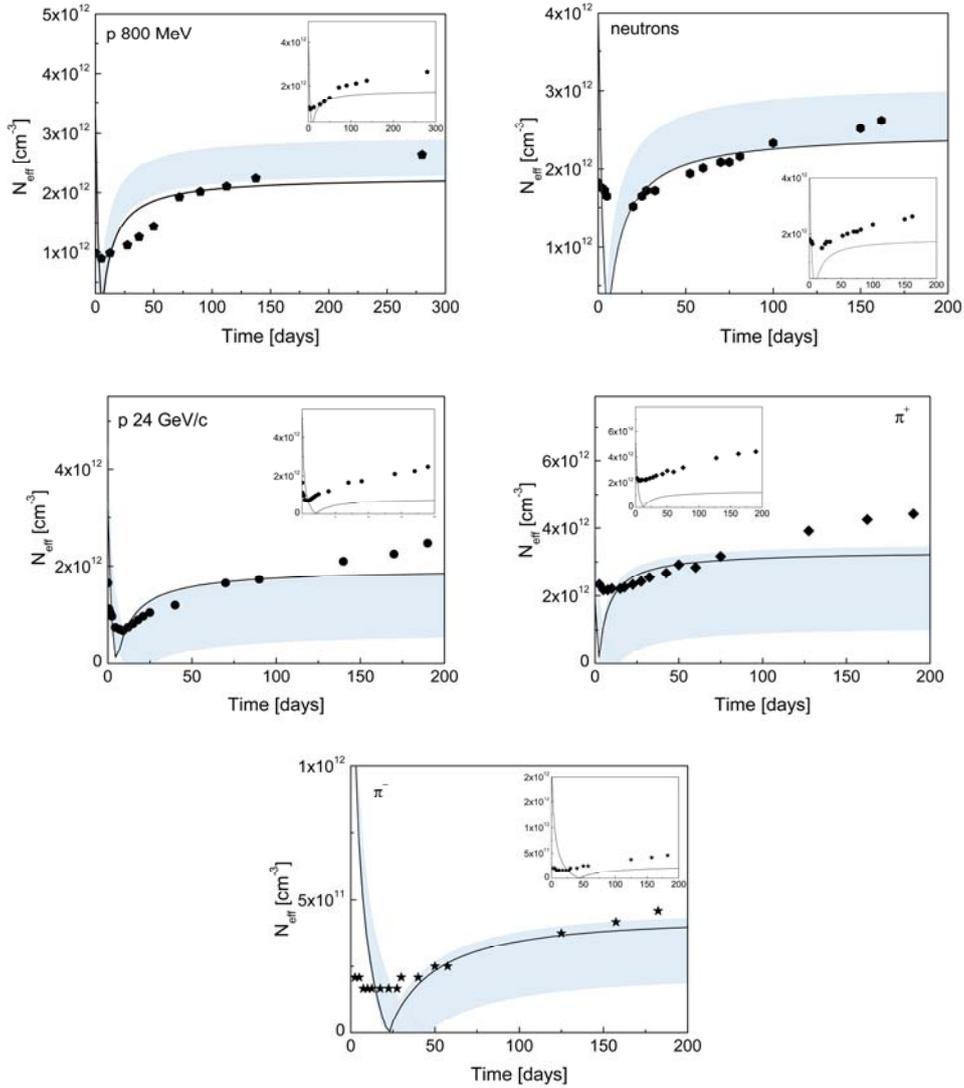

**Figures 3 a ÷ e**
The time dependence of the absolute values of effective doping concentrations after irradiation with: a) neutrons, b) and c) protons of 24 GeV/c and 800 MeV respectively, d) and e) negative and positive pions of 350 MeV/c. In the insert: comparison between experimental data (points) and model calculations without the contribution of the $Si_{FFCD}$ defect (continuous line).

The model, now without free parameters, is able to predict modifications in the behaviour of detectors after irradiations. In reference [35], model calculations have been compared with experimental data on fluence dependence of the effective carrier concentrations of FZ and DOFZ silicon detectors, with different resisitivities (2 kΩcm and 15 kΩcm) and crystal orientations (<100> and <111>), irradiated with 350 MeV/c pions and 24 GeV/c protons. A good agreement with experimental data is obtained, suggesting the robustness of model predictions.

In the comparison of model calculations and experimental data for fluence dependence of the effective carrier concentrations of Cz Si detectors irradiated with pions, the agreement is restricted only to the beginning of irradiation, i.e. to low fluences [35]. In our opinion, this fact could be attributed to the presence of defects associated with oxygen in the material, and consequently supplementary processes they are taking part in during and after irradiation must be added to the reaction scheme.

Thus, considering the contributions of these primary defects to silicon degradation, the old discrepancies between data and model calculations are for the first time solved.

If these conclusions are correct, thus, in conditions of continuous long time irradiation, as e.g. LHC and its upgrades in energy and luminosity, S-LHC and V-LHC respectively, or in the use of the detectors long time in space missions, these contributions will represent a major problem and must be considered.

The model has also been applied to predict the behaviour of FZ and DOFZ Si detectors in the radiation fields estimated for the Large Hadron Collider at CERN, and for its up-grade as Super LHC – see Ref. [35].

## 5. Summary

In this paper, considering the contributions of primary defects as: "classical" vacancy, $Si_{FFCD}$ and self-interstitials to silicon degradation, the old discrepancies between data and model calculations for the leakage current and the effective concentration of charge carriers in the space charge region of detectors exposed to high fluences of particles, disappeared for the first time. The new $Si_{FFCD}$ defect is produced with a concentration of about 10% from all vacancies per act of interaction, and has an energy level in the band gap between $E_c$ – (0.46 ÷ 0.48) eV.
The very good agreement between calculations and experimental data suggests the robustness of model predictions. The contribution of the primary defects to the radiation damage of detectors must be considered in conditions of continuous long time irradiation and/or high fluences

## Acknowledgments

This work was partially supported by the Romanian Scientific National Programmes CERES and MATNANTECH under contracts C4-69/2004 and 219 (404)/2004 respectively.